\documentclass[prb,preprint,showpacs]{revtex4}
\usepackage{graphicx}
\usepackage{amsmath,amssymb,bm}
\begin{document}
\title{Strain and rotation fields of dislocations in graphene }
\author{L.L. Bonilla$^{1,3}$, A. Carpio$^{2,3}$ }
\affiliation {
$^1$G. Mill\'an Institute, Fluid Dynamics, Nanoscience and Industrial
Mathematics, Universidad Carlos III de Madrid, Avda.\ Universidad 30; E-28911 Legan\'es, Spain\\
$^2$Departmento de Matem\'atica Aplicada, Universidad Complutense de Madrid; E-28040 Madrid, Spain\\
$^3$School of Engineering and Applied Sciences, Harvard University, 29 Oxford Street, Cambridge, Massachusetts 02138, USA}
\date{\today}
\begin{abstract}
Strain and rotation fields of dislocations in monolayer graphene have been mapped in a recent experiment. These fields are finite everywhere and differ from those given by linear elasticity 
which does not consider rotation explicitly and predicts infinite rotation and strains at the dislocation point. A derivative regularization of two-dimensional linear elasticity fixes these shortcomings. The theory adds rotation, dislocation and residual strain energies to the usual elastic energy. There are two extra material constants that determine the size of the dislocation core and are determined from experimental data. These findings are useful for studies of dislocations in graphene and for analyzing incipient plasticity in two dimensional crystals. 
\end{abstract}
\maketitle 
\renewcommand{\thefootnote}{\arabic{footnote}}



\section{Introduction} \label{sec:1}

Advanced imaging techniques with high-resolution transmission electron microscopes  have produced accurate images of the displacement and strain fields of dislocations up to 0.03 \AA.\cite{hyt03,zha08} In graphene and other 2D crystals, images of dislocation cores as defects in the crystal lattice and real-time pictures of defect evolution are now being obtained.\cite{mey08,mao11} Gliding and climbing motion of dislocations in graphene has been demonstrated in recent experiments.\cite{war12} In the same experiments, the strain and rotation fields (i.e., the symmetric and the antisymmetric parts of the 2D displacement vector gradient, respectively) near dislocations have been obtained.\cite{war12} These findings have surprising elements that are worth examining and offer the opportunity to understand plastic deformation in nanoscale materials. At the micrometer scale, there are effective computational theories of line dislocations that rely on a mixture of theory and empirical rules for dislocation interaction and motion.\cite{bul06} Precise measurements and theoretical understanding at the nanometer scale could help bridging the gap between scales. On the other hand, controlling electronic properties in graphene through strain engineering is a promising concept\cite{gui10} and dislocations and defects could play an important role. For example, special line defects (grain boundaries) may be used to filter electrons from different graphene valleys (valleytronic devices).\cite{gun11} Experiments have shown that vacancies and point defects produce paramagnetism\cite{lev10} and appropriate strain fields induce strong pseudo-magnetic fields and Landau levels.\cite{nai12} 

Dislocations in two-dimensional (2D) linear elasticity are solutions of the Navier equations with point forces at the dislocation point. In 3D, they correspond to planar edge dislocations that respond to forces supported on the dislocation line.\cite{ll7} Both the strain\cite{hir82} and the rotation are inversely proportional to the distance to the dislocation point and are therefore singular there.\cite{rotation} A simple way to eliminate this unphysical behavior is to regularize elasticity near the dislocation cores by using the correct lattice structure of the crystal. This produces spatially discrete elasticity, an idea that goes back to Frenkel and Kontorova.\cite{FK} Lattice regularization leads naturally to the concept of Peierls stress and of dislocation glide as traveling wave motion;\cite{CB} see Ref.\onlinecite{car08} for the case of graphene. Another widely used regularization is the semi-continuous Peierls-Nabarro (PN) model and its generalizations that contain the interatomic distance $d$ as a parameter.\cite{bul06} The original PN model regularizes the dislocation singularity at the dislocation line but produces unrealistically large strains at the dislocation core (cf. page 223 in Ref. \onlinecite{hir82}). More recently, some authors have replaced the vertical coordinate $y$ (a variable) instead of the interatomic distance (a fixed number) in the PN expressions for $y=0$.\cite{hir82} The resulting formulas,\cite{zha08} including a modification that adds an extra free parameter to the PN expressions,\cite{for51} fit well the experimental strain field in graphene far enough from the dislocation point.\cite{war12} However the modified PN formulas restore the singularity of the strains at the dislocation point, thereby losing the raison d'$\hat{e}$tre of the PN model. No explanation of the rotation caused by dislocations is offered.

In this paper, we present a consistent modification of 2D continuum elasticity that regularizes strains and rotations at the dislocation core and agrees with experimental data in graphene.\cite{war12} See Ref. \onlinecite{laz09} for related work in 3D. The idea is to add three terms to the usual strain energy density in continuum elasticity: (i) a term proportional to the square of the rotation, (ii) a term proportional to the square of the dislocation density vector, and (iii) a background term containing the contribution of the residual stress needed to equilibrate the dislocations. Even if rotating the crystal costs little, the term (i) will forbid the infinite twists at the dislocation point produced by linear elasticity.\cite{rotation} A side effect of this term is that the stress and distortion tensors are not symmetric. The dislocation energy (ii) is not zero wherever there are dislocations and it produces the moment stress needed to balance the antisymmetric part of the stress tensor. The residual stress tensor in (iii) is a sort of {\em bare} stress and the whole theory can be thought of as a derivative regularization\cite{sla71} of continuum elasticity: the equations of motion contain higher derivative terms that produce a smoother {\em dressed} stress. The material constants associated to rotations and dislocations yield two lengths, $l_\Psi$ and $l_\Phi$, that render finite the rotations and the strains at the dislocation point. Over these lengths, rotations and strains differ from the usual expressions of continuum elasticity. We find $l_\Psi$, $l_\Phi$ and the associated material constants by comparing the formulas provided by the theory with experimental data.\cite{war12}  These modifications of linear elasticity in presence of dislocations may pave the way to bridging the nano and micro scales.

\section{Strain energy and equations of motion}\label{sec:2} 
For 2D planar graphene with displacement vector $u_i$ ($u_1=u$, $u_2=v$), the potential energy is split in elastic, rotation, dislocation and background parts
\begin{eqnarray}
 W=
\frac{1}{2}\int [\lambda \beta_{ii}^2 + 2\mu \beta_{(ik)}^2+ 2\gamma\beta_{[ik]}^2+\varsigma \alpha_{i}^2\nonumber\\
-2\sigma_{ik}^0\beta_{ik}]\, dx\, dy,  \label{eq1}
\end{eqnarray}
where $\lambda$ and $\mu$ are the 2D Lam\'e moduli, $\gamma$ is the rotation modulus, $\beta_{(ik)}\equiv (\beta_{ik}+\beta_{ki})/2$ and $\beta_{[12]}\equiv (\beta_{12}-\beta_{21})/2$ are the elastic strain and the rotation, respectively. Sum over repeated indices is implied. The displacement gradients, $
u_{i,j}\equiv\partial_ju_i = \beta_{ij}+\beta^0_{ij}$, 
contain elastic and plastic distortions, $\beta_{ij}$ and $\beta^0_{ij}$, respectively. The dislocation strain energy (with associated modulus $\varsigma\geq 0$) is quadratic in the dislocation density vector $\alpha_i$. The latter can be extracted from the definition of the Burgers vector,\cite{ll7} 
\begin{eqnarray}
-b_i=\oint du_i=\oint dx_j u_{i,j} =\epsilon_{kj}\int \partial_k\beta_{ij} dx\,dy,\label{eq2}
\end{eqnarray}
where we have replaced $\beta_{ij}= u_{i,j}$, the line integral is calculated on a contour encircling the dislocation point, we have used the Stokes theorem, and $\epsilon_{ij}$ is the 2D completely antisymmetric tensor with $\epsilon_{12}=1$. From this expression, the dislocation density vector\cite{ddv} is
\begin{eqnarray}
 \alpha_{i}=\epsilon_{jk}\beta_{ij,k}. \label{eq3}
\end{eqnarray}
 For a point dislocation with Burgers vector $b_i$ located at $x=y=0$, the dislocation density is just $\alpha_i=b_i\delta(x)\delta(y)$. 
%
While $\epsilon_{jk}u_{i,jk}=0$, neither the elastic or the plastic distortions are gradients of functions. In fact, (\ref{eq3}) indicates that the incompatibility of the elastic distortion equals the dislocation density. The last term in (\ref{eq1}) (the residual or background energy) contains the residual stress tensor determined by the equilibrium condition 
\begin{eqnarray}
\sigma_{ij,j}^0=0. \label{eq4}
\end{eqnarray}

The potential energy is a functional of the distortion tensor $\beta_{ij}$. In equilibrium, its variation satisfies
\begin{eqnarray}
\frac{\delta W}{\delta\beta_{ij}(x,y)} = \frac{\partial W}{\partial\beta_{ij}(x,y)}-\partial_k\frac{\partial W}{\partial\beta_{ij,k}(x,y)}=0,  \label{eq5}
\end{eqnarray}
which yields 
\begin{eqnarray}
&& \sigma_{ij}+\varsigma\epsilon_{kj}\alpha_{i,k}=\sigma^0_{ij},  \label{eq6}
\end{eqnarray}
where the stress arises from the first three terms in (\ref{eq1})
\begin{eqnarray}
&& \sigma_{ij}=\lambda \beta_{nn}\delta_{ij} + 2\mu \beta_{(ij)}+ 2\gamma\beta_{[ij]}\nonumber\\
 &&\quad=2\mu\left(\frac{\nu\beta_{nn}}{1-\nu}\delta_{ij} +\beta_{(ij)}\right)+2\gamma\beta_{[ij]}.  \label{eq7}
\end{eqnarray}
Here $\nu=\lambda/(\lambda+2\mu)$ is the 2D Poisson ratio. This implies
\begin{eqnarray} 
&&\beta_{nn}= \frac{1-\nu}{2\mu(1+\nu)}\sigma_{nn}, \quad \beta_{(ij)}=\frac{1}{2\mu}\left(\sigma_{(ij)} -\frac{\nu\sigma_{nn}}{1+\nu}\delta_{ij}\right)\!, \nonumber\\
&& \beta_{[ij]}=\frac{1}{2\gamma}\sigma_{[ij]}.  \label{eq9}
\end{eqnarray}
%
The residual stresses satisfy $\sigma_{ij,j}^0=0$. Then differentiating (\ref{eq6}) and using the compatibility identity, $\epsilon_{kj}\alpha_{i,kj}=0$, we get the usual condition of force equilibrium,
\begin{eqnarray}
 \sigma_{ij,j}=0.  \label{eq8}
\end{eqnarray}

\section{Stress, strain and rotation}\label{sec:3} 
To solve (\ref{eq8}), we write the dislocation density $\alpha_{ij}$ as a functional of the stress $\sigma_{ij}$ and calculate the residual stress for a given dislocation configuration. 
\subsection{Equations of motion for the stress}
Equation (\ref{eq3}) gives the dislocation density in terms of the distortion, which can be calculated as a function of stress using (\ref{eq9}). After some algebra and simplifications through the use of $\sigma_{ij,j}=0$, we obtain
\begin{eqnarray}
&&\alpha_{n,n}= -\frac{1}{2}\left(\frac{1}{\gamma}+\frac{1}{\mu}\right)\Delta\sigma_{[1,2]}, \nonumber\\ 
&&\epsilon_{ki}\alpha_{i,k}=\alpha_{2,1}-\alpha_{1,2}=-\frac{\Delta\sigma_{nn}}{2\mu(1-\nu)},\label{eq12}
\end{eqnarray}
where $\Delta f=\nabla^2f=(\partial^2_1+\partial_2^2)f$. Using 
\begin{eqnarray}
 \epsilon_{k[j}\alpha_{i],k}= \left\{\begin{array}{cc}
 \frac{(-1)^j}{2}\alpha_{n,n},&i\neq j,\\
 0,& i=j,\\
 \end{array} \right. \quad  \epsilon_{k(j}\alpha_{i),k}=
\left\{\begin{array}{cc}-\alpha_{1,2}, &i=j=1,\\
\alpha_{2,1},& i=j=2,\\
\frac{1}{2}(\alpha_{1,1} -\alpha_{2,2}),& i\neq j,\\
\end{array}\right.  \label{eq10}
\end{eqnarray}
that follows from (\ref{eq3}), the antisymmetric part of (\ref{eq6}) becomes
\begin{eqnarray}
\left[1-\frac{\varsigma}{4}\left(\frac{1}{\mu}+\frac{1}{\gamma}\right)\Delta\right]\sigma_{[12]}=\sigma^0_{[12]}.  \label{eq13}
\end{eqnarray}
Similarly, taking the trace in (\ref{eq6}) and using (\ref{eq12}) and (\ref{eq10}), we obtain 
\begin{eqnarray}
\left[1-\frac{\varsigma}{2\mu(1+\nu)}\Delta\right]\sigma_{nn}=\sigma^0_{nn}.  \label{eq14}
\end{eqnarray}

\subsection{Bare and dressed stress functions}
As indicated in Appendix \ref{sec:a1}, Equations (\ref{eq13}) and (\ref{eq14}) can be solved using the Airy stress functions, $\Phi$ and $\Psi$, such that 
%
\begin{eqnarray}
&&\sigma_{11} =\Phi_{,22}-\Psi_{,12},\quad  \sigma_{12}=-\Phi_{,12}+\Psi_{,11},\nonumber\\
&&\sigma_{21} =-\Phi_{,12}-\Psi_{,22},\quad \sigma_{22}=\Phi_{,11}+\Psi_{,12}. \label{eq11}
\end{eqnarray} 
A similar expression holds for the residual stress components, $\sigma^0_{ij}$, in terms  of residual stress functions, $\Phi^0$ and $\Psi^0$. Then we obtain
$\sigma_{nn} =\Delta\Phi$, $\sigma_{[12]} =\frac{1}{2}\Delta\Psi$,
which, inserted into (\ref{eq13}) and (\ref{eq14}), yield the equations
\begin{eqnarray}
&&(1-l_\Phi^2\Delta)\Phi=\Phi^0, \quad (1-l_\Psi^2\Delta)\Psi=\Psi^0,  \label{eq15}\\
&&l_\Phi^2=\frac{\varsigma}{2\mu(1+\nu)},\quad l_\Psi^2=\frac{\varsigma}{4}\left(\frac{1}{\mu}+\frac{1}{\gamma}\right)\!.  \label{eq16}
\end{eqnarray}
Eq. (\ref{eq15}) indicates that $\Phi$ and $\Psi$ are higher derivative regularizations of the usual Airy stress functions. To solve (\ref{eq15}) for a single dislocation located at the origin, we need to first find the residual stress functions $\Phi^0$ and $\Psi^0$. According to Eq.\ (\ref{eq3}), the residual distortion induced by the dislocation satisfies
\begin{eqnarray}
&&\epsilon_{jk}\beta_{ij,k}^0=-b_i\delta(x)\delta(y),\quad \mbox{i.e.,}\nonumber\\ &&  \beta_{i1,2}^0-\beta_{i2,1}^0=-b_i\delta(x)\delta(y). \label{eq17}
\end{eqnarray}
Using (\ref{eq9}) to rewrite (\ref{eq17}) in terms of the residual stresses and stress functions, we obtain
\begin{eqnarray}
&& \Delta^2\Phi^0=-2\mu(1+\nu)(b_1\partial_{2}-b_2\partial_1)\delta(x)\delta(y),
\label{eq18}\\  
&& \Delta^2\Psi^0=\frac{4\gamma\mu}{\mu+\gamma}(b_1\partial_{1}+b_2\partial_2)\delta(x)\delta(y). \label{eq19}
\end{eqnarray}
Thus the residual stress functions are 
\begin{eqnarray}
&& \Phi^0=-2\mu(1+\nu)(b_1\partial_{2}-b_2\partial_1)G(x,y),\nonumber\\ 
&&\Psi^0=\frac{4\gamma\mu}{\mu+\gamma}(b_1\partial_{1}+b_2\partial_2)G(x,y), \label{eq20}\\
&& G(x,y)=\frac{1}{8\pi}r^2\ln r,\quad \Delta^2G(x,y)=\delta(x)\delta(y).\nonumber
\end{eqnarray}
$\Phi^0$ is the stress function of an edge dislocation in 2D linear elasticity.\cite{hir82} The corresponding bare stress, strain and rotation are singular at the origin.
%
For a dislocation with Burgers vector $(b,0)$, the solutions of the inhomogeneous Helmholtz equations (\ref{eq15}) are
\begin{eqnarray}
&&\Phi=-\frac{A}{2} \partial_y\!\!\left\{r^2\ln r+4l_\Phi^2\!\left[\ln r+K_0\!\left(\frac{r}{l_\Phi}\right)\! \right]\!\right\} \!, \label{eq22}\\
&&\Psi=\frac{B}{2} \partial_x\!\!\left\{r^2\ln r+4l_\Psi^2\!\left[\ln r+K_0\!\left(\frac{r}{l_\Psi}\right)\! \right]\!\right\} \!, \label{eq23}
\end{eqnarray}
where $A=\mu b(1+\nu)/(2\pi)$, $B=\mu\gamma b/[\pi(\mu+\gamma)]$. These dressed stress functions become the residual (bare) stress functions (\ref{eq20}) for $l_\Phi=l_\Psi=0$ and coincide with those of a 3D straight edge dislocation\cite{laz09} except for the different value of the prefactor $A$ in (\ref{eq22}). From the stress functions, we can find 
the uniformly bounded stress, strain, rotation and dislocation density vector. 
We get

\begin{eqnarray}
&&\beta_{nn}= -\frac{ b(1+\nu)y}{2\pi r^2}\!\left[1-\frac{r}{l_\Phi}\,K_1\!\left(\frac{r}{l_\Phi}\right)\right]\!,\label{eq24}\\
&&\beta_{[12]} = \frac{\mu bx}{\pi (\mu+\gamma)r^2}\!\left[1-\frac{r}{l_\Psi}\,K_1\!\left(\frac{r}{l_\Psi}\right)\right]\!,\label{eq25}\\
&&\beta_{11}= -\frac{y}{r^4}\left\{\frac{A}{2\mu}\!\left[(1-2\nu)r^2+2x^2+\frac{4l_\Phi^2}{r^2}(y^2-3x^2)\right.\right.\nonumber\\ 
&&\quad\left.-\frac{2(y^2-\nu r^2)r}{l_\Phi}\,K_1\!\left(\frac{r}{l_\Phi}\right) -2(y^2-3x^2)K_2\!\left(\frac{r}{l_\Phi}\right)\right]\nonumber\\
&&\quad-\frac{B}{2\mu}\!\left[x^2-y^2-\frac{4l^2_\Psi}{r^2}(3x^2-y^2)\right.\nonumber\\ 
&&\quad\left.\left.+\frac{2x^2r}{l_\Psi}\,K_1\!\left(\frac{r}{l_\Psi}\right)\!-2(y^2-3x^2)\,K_2\!\left(\frac{r}{l_\Psi}\right)\!\right]\!\right\}\!,\label{eq27}\\
&&\beta_{(12)}= \frac{x}{2\mu r^4}\left\{A\!\left[x^2-y^2-\frac{4l_\Phi^2}{r^2}(x^2-3y^2)\right.\right. \nonumber\\ 
&&\quad\left.-\frac{2y^2r}{l_\Phi}\,K_1\!\left(\frac{r}{l_\Phi}\right)+2(x^2-3y^2)K_2\!\left(\frac{r}{l_\Phi}\right)\right]\nonumber\\ 
&&\quad \left.+B\!\left[y^2-\frac{ r^3}{l_\Psi}\,K_1\!\left(\frac{r}{l_\Psi}\right)\!\right]\!\right\}\!,\quad\beta_{22}=\beta_{nn}-\beta_{11}. \label{eq26}
\end{eqnarray}
\begin{eqnarray}
&&\alpha_{1}= -\frac{b}{2\pi}\,\left\{\frac{1}{4}\Delta^2r^2\ln r+\partial_y^2\left[\ln r+l_\Phi^2\Delta K_0\!\left(\frac{r}{l_\Phi}\right)\right]+\partial_x^2\left[\ln r+l_\Psi^2\Delta K_0\!\left(\frac{r}{l_\Psi}\right)\right]\right\}\nonumber\\
&&= -b(1+l^2_\Phi\partial_y^2+l^2_\Psi\partial_x^2)\delta(x)\delta(y) -\frac{b}{2\pi}\,\left\{l^2_\Phi\partial_y^2\Delta K_0\!\left(\frac{r}{l_\Phi}\right)\!+l^2_\Psi\partial_x^2\Delta K_0\!\left(\frac{r}{l_\Psi}\right)\!\right\}\nonumber\\
&&=-b(1+l^2_\Phi\partial_y^2+l^2_\Psi\partial_x^2)\delta(x)\delta(y)-\frac{b}{8\pi r}\! \left\{\frac{l_\Phi}{r^2}\!\left(6\frac{x^2-y^2}{r^2}-\frac{3r^2}{l^2_\Phi}-2\right)\!K_1\!\left(\frac{r}{l_\Phi}\right)\!\right.\nonumber\\
&&- \left(\frac{3(r^2-4l^2_\Phi)(x^2-y^2)}{4r^3l^2_\Phi}-\frac{3r}{4l^2_\Phi}+\frac{1}{r}\right)\!K_0\!\left(\frac{r}{l_\Phi}\right)\!+\frac{y^2}{2r^3} K_4\!\left(\frac{r}{l_\Phi}\right)-\frac{1}{l_\Phi}K_3\!\left(\frac{r}{l_\Phi}\right)\nonumber\\
&&+\!\left[\!\frac{r^2-3l_\Phi^2}{r^3l^2_\Phi}(y^2-x^2)+\frac{r}{l^2_\Phi}-\frac{1}{r}\!\right]\!K_2\!\left(\frac{r}{l_\Phi}\right)\!+\!\left[\!\frac{r^2-3l^2_\Psi}{r^3l^2_\Psi}(x^2-y^2)+\frac{r}{l^2_\Psi}-\frac{1}{r}\!\right]\!K_2\!\left(\frac{r}{l_\Psi}\right)\nonumber\\
&&+ \left(\frac{3(r^2-4l^2_\Psi)(y^2-x^2)}{4r^3l^2_\Psi}-\frac{3r}{4l^2_\Psi}+\frac{1}{r}\right)\!K_0\!\left(\frac{r}{l_\Psi}\right)\!+\frac{x^2}{2r^3} K_4\!\left(\frac{r}{l_\Psi}\right)-\frac{1}{l_\Phi}K_3\!\left(\frac{r}{l_\Psi}\right)\nonumber\\
&&\left.-\frac{l_\Psi}{r^2}\!\left(6\frac{y^2-x^2}{r^2}-\frac{3r^2}{l^2_\Psi}-2\right)\!K_1\!\left(\frac{r}{l_\Psi}\right)\!\right\}\!, \label{eq48}\\
&&\alpha_{2}=\frac{b}{2\pi}\,\Delta\partial_x\partial_y\left[l^2_\Phi K_0\!\left(\frac{r}{l_\Phi}\right)-l^2_\Psi K_0\!\left(\frac{r}{l_\Psi}\right)\right]\nonumber\\
&&=\frac{bxy}{4\pi r^2}\left\{\!\left[\frac{r^2-3l_\Phi^2}{r^2l_\Phi^2}K_2\!\left(\frac{r}{l_\Phi}\right)\!-\frac{r^2-3l_\Psi^2}{r^2l_\Psi^2} K_2\!\left(\frac{r}{l_\Psi}\right)\!\right]\!+\frac{1}{4l_\Phi^2}K_4\!\left(\frac{r}{l_\Phi}\right)\!-\frac{1}{4l_\Psi^2} K_4\!\left(\frac{r}{l_\Psi}\right)\right.\nonumber\\
&&\left.+\frac{3(r^2-4l_\Phi^2)}{4r^2l_\Phi^2}K_0\!\left(\frac{r}{l_\Phi}\right)\!-\frac{3(r^2-4l_\Psi^2)}{4r^2l_\Psi^2}K_0\!\left(\frac{r}{l_\Psi}\right)\!-\frac{6l_\Phi}{r^3} K_1\!\left(\frac{r}{l_\Phi}\right)\! +\frac{6l_\Psi}{r^3}K_1\!\left(\frac{r}{l_\Psi}\right)\!\right\}\!.
\label{eq49}
\end{eqnarray}
We can calculate the Burgers vector distribution by integrating the dislocation density on a disk of radius $r$ centered at the origin. The resulting vector is $(b(r),0)$ with 
\begin{eqnarray}
b(r)&=& -\int_0^r\int_0^{2\pi}\alpha_1 r\, dr\,d\theta= b+\frac{b}{2}\, r\partial_r\!\left(\partial_r^2+\frac{1}{r}\partial_r\right)\!\!\left[l^2_\Phi K_0\!\left(\frac{r}{l_\Phi}\right)\!+l^2_\Psi K_0\!\left(\frac{r}{l_\Psi}\right)\!\right]\!\nonumber\\
&=& b+\frac{b}{4}\!\left[K_0\!\left(\frac{r}{l_\Phi}\right)\! +K_2\!\left(\frac{r}{l_\Phi}\right)\!+ K_0\!\left(\frac{r}{l_\Psi}\right)\! +K_2\!\left(\frac{r}{l_\Psi}\right)\!+ \!\left(\frac{2l_\Phi}{r} - \frac{3r}{2l_\Phi}\right)\!K_1\!\left(\frac{r}{l_\Phi}\right)\! \right.\nonumber\\
&+&\left. \!\left(\frac{2l_\Psi}{r} - \frac{3r}{2l_\Psi}\right)\!K_1\!\left(\frac{r}{l_\Psi}\right)\! - \frac{r}{2l_\Phi}K_3\!\left(\frac{r}{l_\Phi}\right)\!- \frac{r}{2l_\Psi}K_3\!\left(\frac{r}{l_\Psi}\right)\!\right]\!.\label{eq28}
\end{eqnarray}
The distribution $b(r)$ monotonically increases from 0 to $b$ as $r\to \infty$.

\begin{figure}
\begin{center}
\includegraphics[width=8cm]{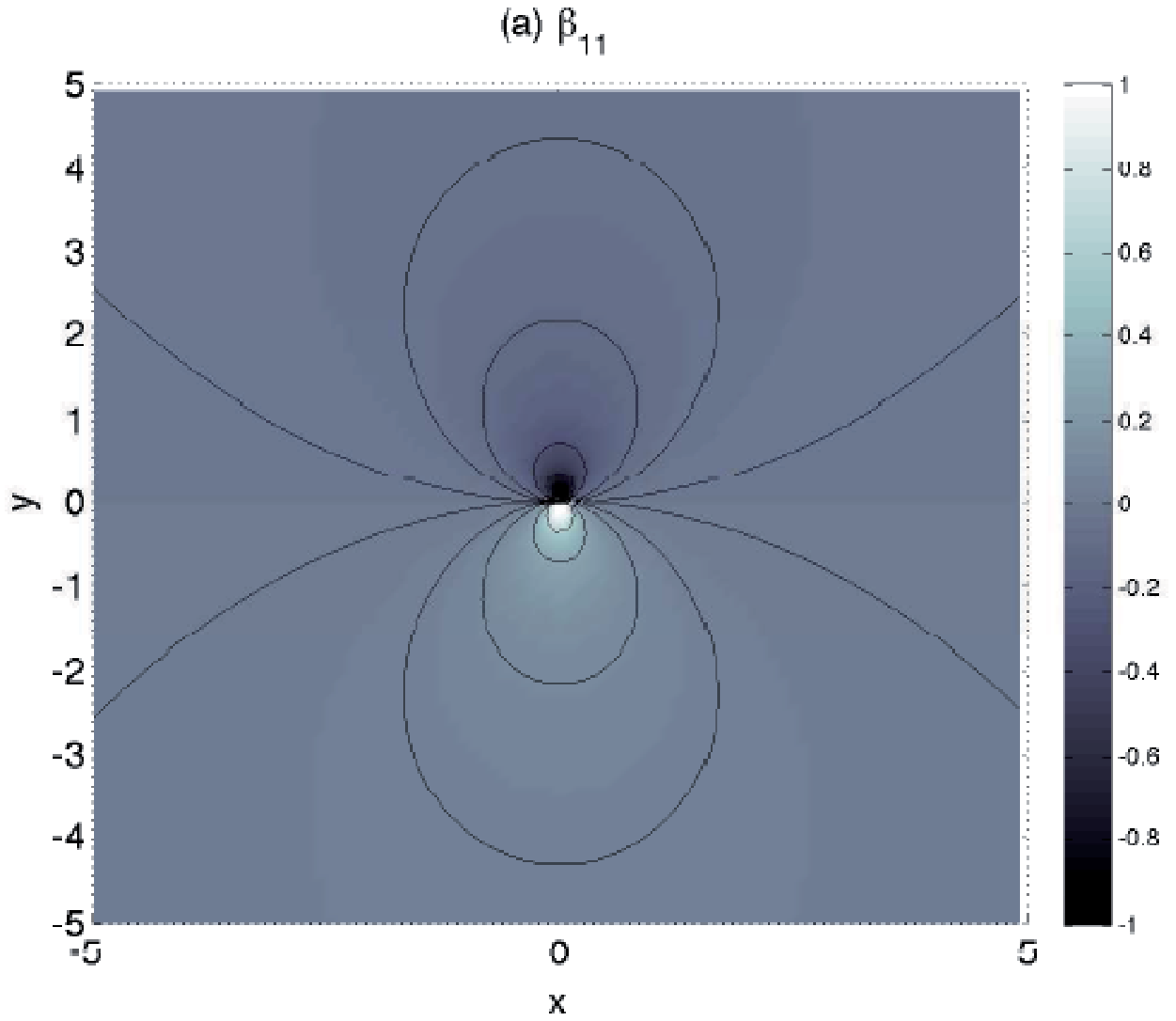}
\includegraphics[width=8cm]{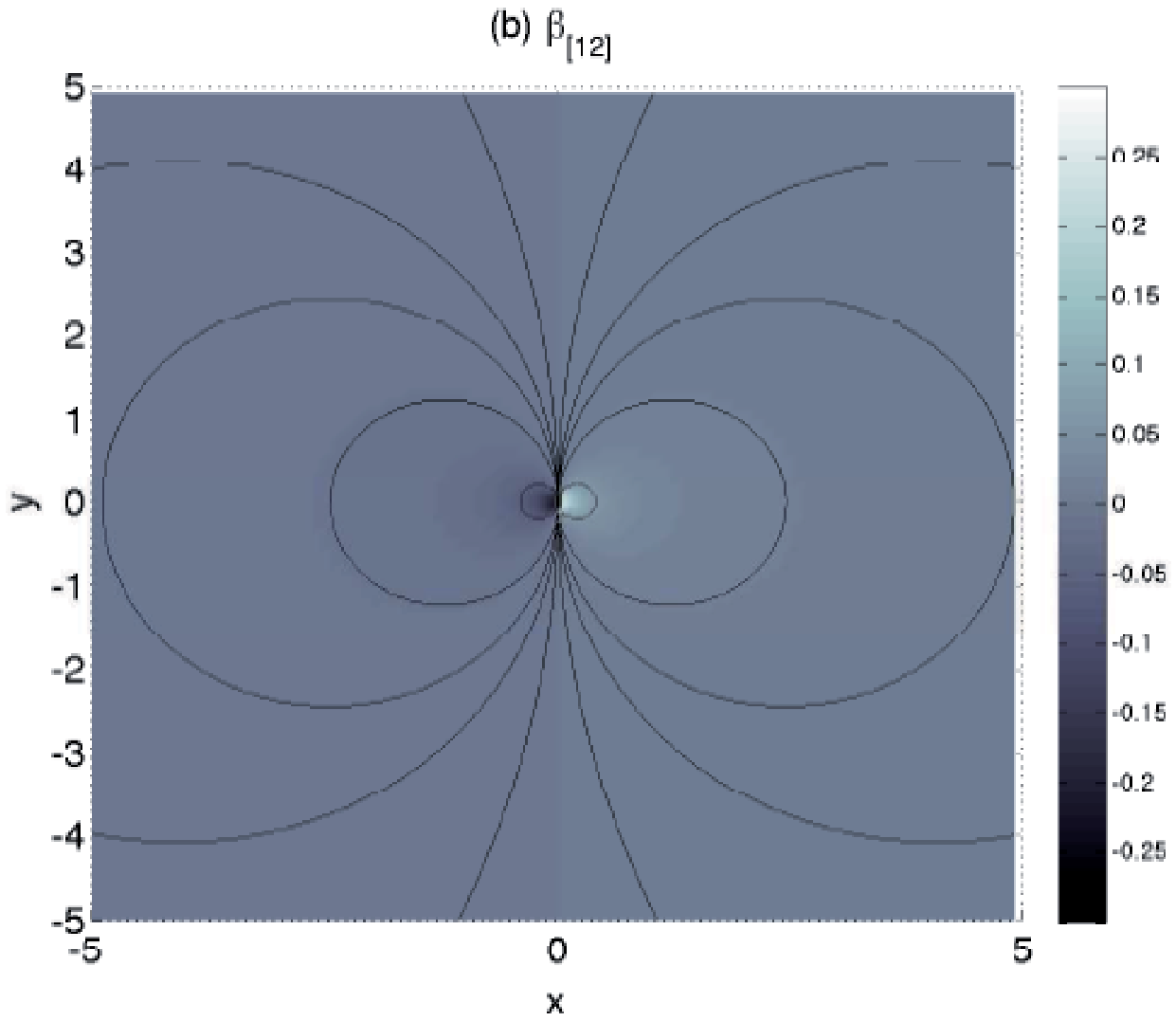}
\caption{Contours of (a) strain $\beta_{11}$, (b) 
rotation $\beta_{[12]}$. }
\label{fig1}
\end{center}
\end{figure}
\begin{figure}
\begin{center}
\includegraphics[width=8cm]{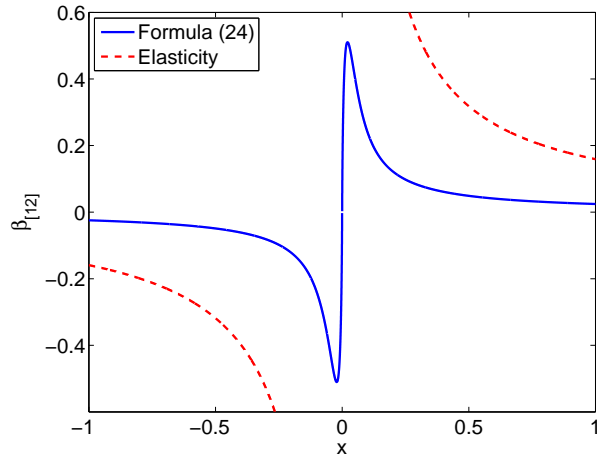}
\caption{(Color online) Rotation profiles at $y=0$. Solid line corresponds to (\ref{eq25}) and dashed line to continuum elasticity, $\beta_{[12]}=b\cos\theta/(2\pi r)$. }
\label{fig2}
\end{center}
\end{figure}

\section{Results}\label{sec:4} 
Figure \ref{fig1} shows the contours of the strain $\beta_{11}(x,y)$ and the rotation $\beta_{[12]}(x,y)$. Fig. \ref{fig2} compares the rotation along the horizontal axis given by (\ref{eq25}) and the rotation given by linear elasticity, which is unbounded as $x\to 0$. We have found the moduli $\gamma$ and $\varsigma$ (equivalently, $l_\Phi$ and $l_\Psi$) by fit to experiments.\cite{war12} The strain $\varepsilon_{xx}$ inspired by the PN model and Ref. \onlinecite{for51} is\cite{zha08}
\begin{eqnarray}
&&\varepsilon_{xx}=-b(1-\nu)\sin\theta\nonumber\\
&&\quad\times\frac{4(1-\nu)^2\cos^2\theta+(2a_f-1)a_f^2\sin^2\theta}{\pi r\, [4(1-\nu)^2\cos^2\theta+a_f^2\sin^2\theta]^2},  \label{eq}
\end{eqnarray}
when written in polar coordinates. See also Supplementary Eq. (2) in Ref.~\onlinecite{war12}. Note that the strain (\ref{eq}) is singular at $r=0$, thereby defeating the purpose of the PN model. The best fit to graphene data at different nonzero distances $r$ is reported to be $a_f=1.5$.\cite{war12} The contours of this strain and those of $\beta_{11}$ given by (\ref{eq27}) should agree. On the other hand, the rotation 
about a dislocation (heptagon-pentagon defect) in graphene goes from -0.5 to 0.5 radians.\cite{war12} Using 0.5 radians for the maximum rotation in (\ref{eq25}) fixes the scale $l_\Psi$. We find that $\varsigma/(\lambda+2\mu)=6 \times 10^{-4}b^2$ and $\gamma=12\mu$. The Burgers vector of a dislocation in graphene is the lattice constant, $b=1.42\sqrt{3}$ \AA. At 300 K, we have $\mu=9.95$ eV/\AA$^2$, $\lambda+2\mu=22.47$ eV/\AA$^2$,\cite{zak09} and therefore $\varsigma= 0.08155$ eV and $\gamma=119.4$ eV/\AA$^2$. Furthermore the characteristic lengths $l_\Phi=0.025 b$ and $l_\psi=0.019b$ are below 0.1 \AA. The dressed strain and rotation behave similarly to the bare ones a short distance away from the dislocation point. At larger distances, the empirical formula (\ref{eq}) with $a_f=1.5$ used to fit our theory gives similar strain contours.

We can compare the results given by the present theory with those produced by periodized discrete elasticity.\cite{car08} For this purpose, we have approximated the partial derivatives in the definition of the distortion tensor by finite differences using the honeycomb lattice. Very few values are available to reconstruct partial derivatives from finite differences at the core of the dislocation which would produce a pixelated figure (not shown). The resulting density plot is similar to that yielded by our regularized continuum model, with positive (negative) values of rotation for $x>0$ ($x<0$). In the case of periodized discrete elasticity, the symmetry $y\to-y$ of Fig. \ref{fig1}(b) of the continuum theory is broken due to discreteness effects which are very strong near the dislocation point. This symmetry breakdown is also visible in experiments; see Supplementary Figures S10, S11 and S15 in Ref.\onlinecite{war12}.

\section{Conclusions} \label{sec:6}
We have calculated the strain and rotation contours about a single dislocation in graphene by regularizing continuum elasticity in two ways. A lattice regularization (periodized discrete elasticity)\cite{car08} eliminates the singularities at the dislocation point and gives contour plots that agree with experimental data.\cite{war12} We can also carry out a {\em derivative regularization} of elasticity by means of a continuum theory that introduces rotation, dislocation and background (residual) strain densities. This theory contains two additional material constants that we have estimated by using data from experiments. While we have considered a planar graphene sheet, off-plane vertical displacements can be incorporated to our continuum theory by including a bending energy and using ideas similar to those in Ref. \onlinecite{che11}. 

\acknowledgments
This work has been supported by the Spanish Ministerio de Econom\'\i a y Competitividad grants FIS2011-28838-C02-01,  FIS2011-28838-C02-02 and FIS2010-22438-E (Spanish National Network Physics of Out-of-Equilibrium Systems). The authors thank M.P. Brenner for hospitality during a stay at Harvard University financed by Fundaci\'on Caja Madrid mobility grants.

\appendix
\setcounter{equation}{0}
\renewcommand{\theequation}{A.\arabic{equation}}
\section{Proof of the existence of the Airy stress functions}
\label{sec:a1}
The arguments in this Appendix are adapted from Ref. \onlinecite{min63}. Since $u_{i,j}=\beta_{ij}+\beta^0_{ij}$, (\ref{eq3}) implies that $\alpha_i=-\alpha_i^0$, where $\alpha_i^0$ is defined by replacing the residual distortion $\beta_{ij}^0$ instead of the distortion in (\ref{eq3}). In (\ref{eq6}), let us denote 
\begin{eqnarray}
\sigma_{ij}^\psi=\frac{\varsigma}{2} \epsilon_{kj}\alpha_{i,k},
\quad \sigma_{ij}^{\psi 0}=\frac{\varsigma}{2} \epsilon_{kj}\alpha_{i,k}^0\Longrightarrow (\sigma_{ij}+\sigma_{ij}^\psi)-(\sigma_{ij}^0+\sigma_{ij}^{\psi 0})=0.\label{c1}
\end{eqnarray}
Clearly,
\begin{eqnarray}
\sigma_{ij,j}=0, \quad\sigma_{ij,j}^\psi=0,\quad\sigma_{ij,j}^0=0, \quad\sigma_{ij,j}^{\psi 0}=0,  \label{c2}
\end{eqnarray}
and we can introduce eight functions $\phi$, $\tilde{\phi}$, $\psi$, $\tilde\psi$, $\phi^0$, $\tilde{\phi}^0$, $\psi^0$, $\tilde{\psi}^0$,
\begin{eqnarray}
\sigma_{11}=\phi_{y}, \quad\sigma_{12}=-\phi_{x}, \quad \sigma_{21}=-\tilde{\phi}_{y}, \quad\sigma_{22}=\tilde{\phi}_{x},\label{c3}\\
\sigma_{11}^\psi=\psi_{y}, \quad\sigma_{12}^\psi=-\psi_{x}, \quad \sigma_{21}^\psi=-\tilde{\psi}_{y}, \quad\sigma_{22}^\psi=\tilde{\psi}_{x},  \label{c4}
\end{eqnarray}
with similar definition for the residual stresses. In this Appendix, subscripts in the stress functions imply partial derivatives: $\phi_x=\partial_x\phi$, etc. To show that these four functions can be written in terms of only four stress functions, we proceed as follows. Firstly, we show that $\psi$ and $\tilde{\psi}$ are proportional to the components of the dislocation density vector. $\psi_y=\sigma^\psi_{11}=-\varsigma\alpha_{1,2}$ and $-\psi_x=\sigma^\psi_{11}=\varsigma\alpha_{1,1}$ imply that $\psi=-\varsigma\alpha_1$. Similarly, we can show that $\tilde{\psi}=\varsigma\alpha_2$, $\tilde{\psi}^0=\varsigma\alpha_2^0$, $\psi^0=-\varsigma\alpha_1^0$.

Now the antisymmetric part of (\ref{eq6}) can be written as $\sigma_{[12]}+\frac{1}{2} \varsigma\epsilon_{k[2}\alpha_{1],k}-(\sigma_{[12]}^0+\frac{1}{2} \varsigma\epsilon_{k[2}\alpha_{1],k}^0) =0$, i.e.
\begin{eqnarray}
\sigma_{12}-\sigma_{21}+\frac{1}{2} \varsigma(\alpha_{1,1}+\alpha_{2,2})-\sigma_{12}^0+\sigma_{21}^0-\frac{1}{2} \varsigma(\alpha_{1,1}^0+\alpha_{2,2}^0)=0. \label{c5}
\end{eqnarray}
Using (\ref{c3}), this equation becomes
\begin{eqnarray}
0=\partial_y\!\left(\tilde{\phi}+\frac{\varsigma}{2}\alpha_2\right)\!-\partial_x\!\left(\phi-\frac{\varsigma}{2}\varsigma\alpha_1\right)\!-\partial_y\!\left(\tilde{\phi}^0+\frac{\varsigma}{2}\alpha_2^0\right)\!+\partial_x\!\left(\phi^0-\frac{\varsigma}{2}\varsigma\alpha^0_1\right)\!. \label{c6}
\end{eqnarray}
From which we can set
\begin{eqnarray}
&&\tilde{\phi}+\frac{1}{2}\varsigma\alpha_2=\Phi_x,\quad \phi-\frac{1}{2}\varsigma\alpha_{1}= \Phi_y,\quad \tilde{\phi}^0+\frac{1}{2}\varsigma\alpha_2^0=\Phi_x^0,\quad \phi^0-\frac{1}{2}\varsigma\alpha_{1}^0= \Phi_y^0 \Longrightarrow\nonumber\\
&& \tilde{\phi}=\Phi_x-\frac{\varsigma}{2}\alpha_2,\quad\phi=\Phi_y+\frac{\varsigma}{2}\alpha_1,\quad\tilde{\phi}^0=\Phi_x^0-\frac{\varsigma}{2}\alpha_2^0,\quad\phi^0=\Phi_y^0+\frac{\varsigma}{2}\alpha_1^0, \label{c7}
\end{eqnarray}
in terms of new functions $\Phi$ and $\Phi^0$. The strain compatibility conditions $\epsilon_{ij}\alpha_{i,j}= 0$, $\epsilon_{ij}\alpha_{i,j}^0= 0$, imply that the dislocation density vector derives from a potential, and we have $\varsigma\alpha_1=-\Psi_x$, $\varsigma\alpha_2=-\Psi_y$ and similar expressions for $\alpha_i^0$. Then (\ref{c7}) becomes
\begin{eqnarray}
 \tilde{\phi}=\Phi_x+\Psi_y,\quad\phi=\Phi_y-\Psi_x, \quad\psi=\Psi_x,\quad \tilde{\psi}=-\Psi_y,  \label{c8}
\end{eqnarray}
with similar expressions for the residual stress functions. Substituting this into (\ref{c3}) and (\ref{c4}), we obtain
\begin{eqnarray}
&&\sigma_{11}=\Phi_{yy}-\Psi_{xy}, \quad\sigma_{12}=-\Phi_{xy}+\Psi_{xx}, \quad \sigma_{21}=-\Phi_{xy}-\Psi_{yy}, \quad\sigma_{22}=\Phi_{xx}+\Psi_{xy},\label{c9}\\
&&\sigma_{11}^\psi=\Psi_{xy}, \quad\sigma_{12}^\psi=-\Psi_{xx}, \quad \sigma_{21}^\psi=\Psi_{yy}, \quad\sigma_{22}^\psi=-\Psi_{xy}, \label{c10}
\end{eqnarray}
with similar expressions for the residual stress components and residual stress functions. Equation (\ref{c9}) is the same as (\ref{eq11}). If $\Psi=0$, (\ref{c9}) is the usual relation between the stresses and the Airy stress function of linear elasticity.


\begin{thebibliography}{28}
\bibitem{rotation}
For a 3D straight edge dislocation, the displacement vector $(u,v)$ of linear elasticity given in References \onlinecite{hir82} or \onlinecite{ll7} yields a rotation $\beta_{[12]}=(\partial_yu-\partial_xv)/2=b\cos\theta/(2\pi r)$, where $r=\sqrt{x^2+y^2}$ and $\theta=\tan^{-1}(y/x)$ are polar coordinates and $(b,0)$ is the Burgers vector. For a 2D dislocation, we obtain the same formula.
\bibitem{ddv}
In 3D the dislocation density is a rank two tensor, not a vector, and the theory is more complicated.\cite{laz09}
\end{thebibliography}
\end{document}